\documentclass[pre,twocolumn,showpacs]{revtex4-1}

\usepackage{graphicx}
\usepackage{amsmath}
\usepackage[dvips]{color}

\newcommand{\bnabla}{\mbox{\boldmath $\nabla$}}

\newcommand{\kt} {k_{\rm B}T}

\newcommand{\bea}{\begin{eqnarray}}
\newcommand{\eea}{\end{eqnarray}}
\newcommand{\vect}[1]{\mathbf{#1}}

\begin{document}

\title{Extended Poisson--Boltzmann descriptions of the electrostatic double
  layer: implications for charged particles at interfaces}

\author{Derek Frydel$^{a}$ and Martin Oettel$^{b}$ \\
\textnormal{$^{a}$ Universidade Federal do Rio Grande do Sul, Porto Alegre, Brazil\\
$^{b}$ Institut f\"ur Theoretische Physik II, Heinrich Heine Universit\"at D\"usseldorf, Germany and 
        Institut f\"ur Angewandte Physik, Eberhard Karls Universit\"at T\"ubingen, Germany \\
martin.oettel@uni-tuebingen.de, dfrydel@gmail.com}}


\date{\today}


\pacs{
}

\maketitle

\section{Introduction}\label{sec_oettel.1}

Whenever a charged surface is immersed in a thermal solvent with mobile co-- and counterions
(usually an aqueous electrolyte),
a screening layer of counterions will form in the vicinity of the charged surface, giving rise
to the so called ``double layer''. Systematic investigations of the double layer physics and chemistry 
have been conducted for more than one hundred years and have been motivated by various practical
questions such as the behavior of battery electrodes or the stability of charged colloid solutions.   
Very often surfaces can be highly charged (e.g. colloid surfaces attain charge densities of
about one elementary charge $e$ per nm$^2$) such that the electrostatic potential $\psi$ near the 
surface exceeds the equivalent of the thermal energy $\kt=1/\beta$ by far (for room temperature $\kt/e$ corresponds
to a potential of 25 mV). Thus the arrangements of water molecules and counterions near and at the surface
should display fairly strong correlations, resulting in e.g. nontrivial potential and dielectric profiles.

Such correlations are neglected in the ``work horse'' model of charges in solutions, the Poisson--Boltzmann theory.
Here, Poisson's equation $\bnabla(\epsilon\bnabla \psi(\vect r))= -e(c_+(\vect r)-c_-(\vect r))$ is coupled with
a strongly simplifying assumption on the density distribution of positive and negative ions,
$c_{\pm} = c_s\exp( \mp e \beta \psi(\vect r))$, i.e a Boltzmann distribution of charged point particles
(having no correlations) subject to the electrostatic potential (the ions are monovalent for simplicity and
$\epsilon$ is the dielectric constant of the solvent, $c_s$ is the bulk number density of positive/negative 
ions in the solvent). The resulting Poisson--Boltzmann equation
for the electrostatic potential 
\bea
 \label{eq:pb}
  \bnabla \cdot(\epsilon \bnabla \psi) = 2 c_s e \sinh(\beta e \psi)  \;.
\eea
still poses a formidable nonlinear problem, analytically solvable only in special circumstances such as the geometry
of a charged wall.
For $\epsilon={\rm const.}$, it is convenient to introduce the Debye--H\"uckel screening length via $\kappa^{-1}= (\epsilon/(2 c_s \beta e^2))^{1/2}$ 
and the dimensionless potential $\phi=\beta e\psi$. Using these quantities, the Poisson--Boltzmann equation
becomes {$\bnabla \cdot \bnabla \phi = \kappa^2 \sinh\phi$}, and the significance of the screening length becomes immediately obvious
in its linearized form, {$(\bnabla \cdot \bnabla - \kappa^2) \phi=0$}, which entails an exponentially decaying potential with
characteristic length $\kappa^{-1}$ around a localized charge distribution.

Very often charged particles in electrolytes are characterized using properties of the linearized Poisson--Boltzmann
equation. The {\em effective} or {\em renormalized charge} or {\em charge density} is the one appropriate for the charged particle
immersed in a hypothetical ``linear Poisson--Boltzmann" medium which exhibits the same asypmptotic potential decay as
the ``real" charged particle in the ``real" (and complicated) electrolyte. If two particles immersed in the electrolyte are 
sufficiently far apart,
then their interaction energy is proportional to the product of their effective charges -- this illustrates that the charge
renormalization concept is very useful in characterizing charged colloid suspensions. However, the effective charge density 
of particles in bulk electrolytes contains very little information about the precise structure and potential profile in the 
double layer. This is best {illustrated} by the well--known solution for a charged wall immersed in an electrolyte in 
Poisson--Boltzmann theory.  Using dimensionless
charge densities given by $\tilde \sigma = \beta e/(\kappa \epsilon)\;\sigma$
the effective charge density $\tilde \sigma^{\rm eff}$ of the wall is given in terms of the
bare charge density $\tilde\sigma_{c}$ by
\bea
 \label{eq:pbwall_ren}
 \tilde \sigma^{\rm eff} = 4 \left( \sqrt{1+\frac{4}{\tilde\sigma_{c}^2}} - \frac{2}{\tilde\sigma_{c}^2} \right)\;.
\eea
For colloidal matter, $\sigma_{c} \lesssim 1$ $e$/nm$^2$ and except for situations
where $\kappa^{-1} \lesssim 0.3$ nm (or $c_s \gtrsim 1$ M) we are in the saturated regime
where $\tilde \sigma^{\rm eff}=4$. We can draw an important conclusion from this result: Although we might expect the Poisson--Boltzmann
description to fail near the charged particles for typical colloid charge densities, we must expect that the dimensionless effective
charge density is {\em roughly constant for modest variations of $\sigma_c$ and $\kappa^{-1}$} and its determination will not allow
to discriminate between different models of the double layer.  
Thus by measuring pair interactions between charged particles we mostly learn which area $A$ of the particle's surface is covered with charge
(since $q^{\rm eff}=A\sigma^{\rm eff}$) and 
we do not have access to, say,  the potential at the particle surface.

\section{Charged particles at electrolyte interfaces}

\begin{figure}[tbh]
\vspace{0.6cm}
\centerline{\resizebox{0.45\textwidth}{!}
{\includegraphics{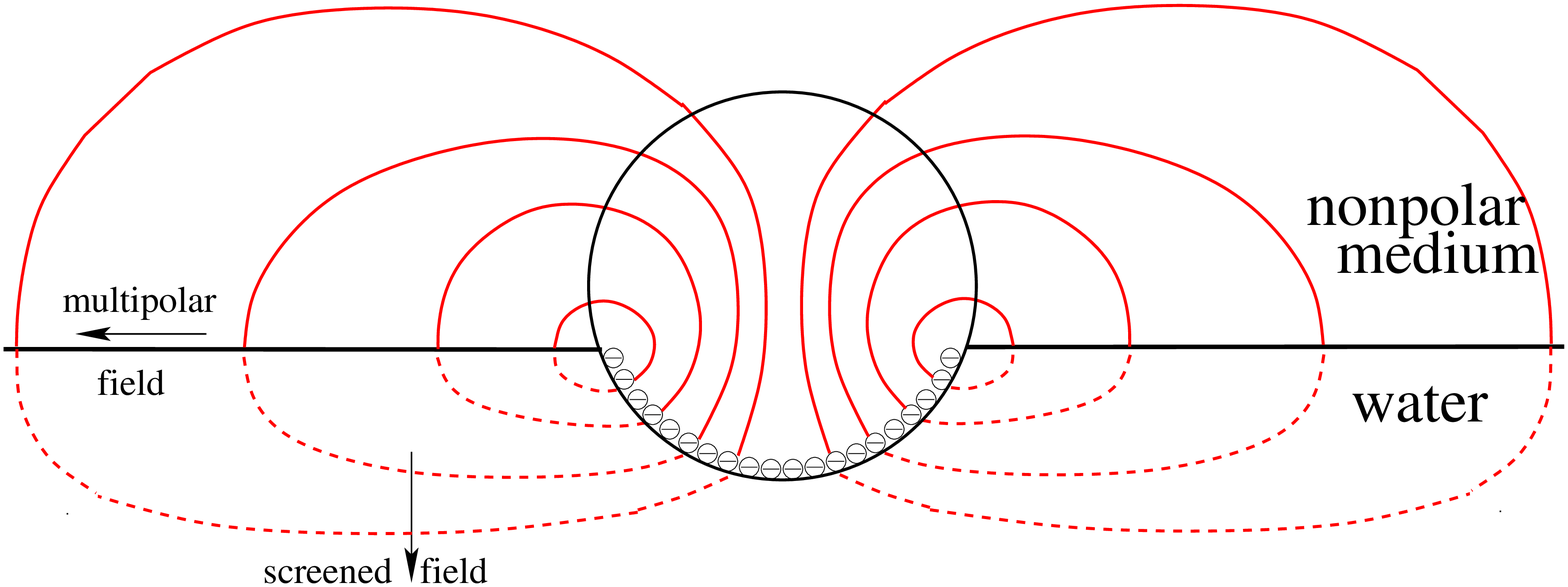}}}
\caption{Sketch of electric field lines around a charged colloid located at
  the interface between water and a nonpolar medium. The asymptotics
  of the electrostatic potential and the associated electric field at the interface
  is mainly determined by field lines
  originating from the colloid charges and passing through the colloid and the nonpolar
  medium. Field lines through the water (containing ions) are screened 
 (indicated by the dashing of the lines).}
\label{fig:fieldlines}
\end{figure}

The situation is different for charged particles at the interface between an electrolyte and a nonpolar medium (air, oil),  
see Fig.~\ref{fig:fieldlines}. If the dielectric constant of the nonpolar medium is low (${\epsilon/\epsilon_0} \lesssim 2$,
$\epsilon_0$ is the dielectric constant of vacuum) then
surface charges can only reside on the colloid surface exposed to the electrolyte. The double layer screens the colloid 
charge only to the extent that no monopole field is present along the interface. However, higher multipole fields
are present, and the leading multipole is a dipole with moment $p_z$ oriented perpendicular to the interface (regardless of the
charge distribution on the colloid \cite{Dom08}). Consequently, the electrostatic interaction energy between two charged 
colloids at the interface with mutual distance $d$ is long--ranged and $\propto p_z^2/d^3$.  
If the local curvature at the colloid surface is smaller than $\kappa$ then the double layer can be approximated by the
double layer of a charged wall in the vicinity of each point at the colloid--electrolyte interface. 
The dipole moment $p_z=A^{\rm eff} p'_{\rm wall}$ can be calculated from the aerial dipole moment density 
$p'_{\rm wall}$ and an effective area $A^{\rm eff}$ which depends on the precise colloid geometry. 
The crucial observation is that $p'_{\rm wall} = \epsilon_0 \psi_{\rm wall}$ is proportional to the
contact potential at the wall.\footnote{Let the wall be located at $z=0$ and the double layer be located in the half--space $z>0$. 
Then $p'_{\rm wall} = \int_{0^+}^\infty (z\rho(z)+P(z)) dz$ where $\rho(z)$ is the charge density in the double layer
and $P(z)=-(\epsilon(z)-\epsilon_0)\partial_z\psi(z)$ is the polarization density of the solvent.  
Since $\partial_z(\epsilon\partial_z \psi) = \rho$, we have $p'_{\rm wall} = \int_{0^+}^\infty \epsilon_0 \partial_z \psi
= \epsilon_0 \psi_{\rm wall}$.} 
Thus the interaction energy at large distances between two colloids is proportional to the square of the
wall contact potential, and such a dependence is very much different from the dependence on the
effective (saturated) charges for the interaction between colloids in bulk electrolyte \cite{Fry07}. 
The Poisson--Boltzmann equation gives $\beta e p'_{\rm wall} \approx 2\epsilon_0 \ln \tilde\sigma_c$
for $\tilde\sigma_c \gg 1$, so already here is seen that the dipole moment will not saturate
with increasing charge density on the colloid and depends logarithmically on both screening length and
charge density, $p_z \propto \ln(\kappa^{-1}\sigma_c) + {\rm const.}$ (see the definition of $\tilde\sigma$).

{\em Experiments. --} The dipole moment can be calculated from the interaction potential between charged colloids
and this can be inferred through laser tweezers on isolated pairs of colloids, from inversion of measured pair
correlations or from elasticity measurements on 2d crystals at the interface. Whereas the logarithmic dependence
on $\kappa^{-1}$ has received much support from laser tweezer data \cite{Par08}, the absolute magnitude
of $p_z$ from the Poisson--Boltzmann equation is too small by about a factor of 5 compared to results
of careful measurements using all three routes \cite{Mas10}.  This points to a severe underestimation of the
contact potential at a charged wall in the Poisson--Boltzmann equation.

\section{Modification strategies for improving the Poisson--Boltzmann description}
\label{sec:mod_PB}

The deficiencies of and possible amendments to the Poisson--Boltzmann equation (\ref{eq:pb}) have been discussed for a long time, 
starting with the work of Bikerman \cite{Bik42}.  {Recently, starting with Ref.~\cite{Bor97}, the modified Poisson-Boltzmann approach 
received renewed interest.} We are interested in modifications which preserve the type of equation as a differential equation 
for the potential $\psi$.  
{The electrostatic interactions thus remain on the mean-field level, and the modifications introduce contributions that do not 
stem from correlations.  This approach is suitable to investigate {certain aspects of} ``ion specificity'' in the weak-coupling limit.  }

\begin{enumerate}

 \item {\em Finite size effects.} -- A finite size of the co-- and counterions (e.g. {through a} hard--sphere volume $v$) will 
  inevitably lead to layering
  effects in the density profile of the ions near the charged surface. {In aqueous solution, 
  the effective ion size is increased as a result of hydration. Thus, in a strict sense}, the finite ion size contributions require
  nonlocal treatment,
  however, one can {capture these effects using local corrections, for example, by}
  taking into account a {local} change in the ion chemical potential $\mu_{\pm}$ (or ion osmotic pressure $p$) through local density changes.
  In the Poisson--Boltzmann equation ions are treated as ideal, thus ${\beta}\mu_\pm = \ln {v}c_\pm$ {(or $\beta p=\rho$)}. A popular
  modification (MPB, modified Poisson--Boltzmann equation) \cite{Bor97} has been derived by lattice gas considerations, leading to
  ${\beta}\mu_\pm = \ln {v}c_\pm - \ln(1-vc_+ - v c_-)$ (or ${\beta}p=-v^{-1}\ln(1-vc_+ - v c_-)$), {where the excess chemical 
  potential, $v^{-1}\ln(1-vc_+ - v c_-)$, accounts for the repulsive (excluded volume) ionic interactions.}
  {{Such a} local correction places an upper bound on the local density, leading to density "saturation" in the region of high ionic
  concentration. Far from the concentrated region, the density is merely shifted by some length $d$, $c^{\rm MPB}(x) = c^{\rm PB}(x-d)$, 
  in relation to the Poisson--Boltzmann result. Thus, the screening behavior far from the concentrated region is not exactly altered, 
  but rather the effective position of a charged surface is shifted into {the solution}. Roughly, the changes associated with the finite 
  size effects can be quantified with an additional length scale $v\sigma_c$, which is the size of a layer of counterions that would form 
  if all counterions were forced into close-packed configuration. The extent of the finite size effects is estimated as the ratio of this 
  length with other relevant lengths, such as the screening or Gouy-Chapman length. }
  Although the lattice gas modification {corresponds to a not very precise equation of state for hard spheres, 
  application of MPB to inhomogeneous ionic solution} yields predictions that compare rather well with exact results. 
  This {points} to some internal cancellation of errors. 

 \item {\em Dielectric inhomogeneity of the solvent.} --
  {The solvent medium, made up of water molecules {with permanent dipoles}, acts as a dielectric {which} is polarized in an external 
  electrostatic field and {thus screens the electrostatic potential within the medium}.
  Thermal fluctuations favor isotropic orientation of dipoles, and the electrostatic 
  interactions favor alignment with the field lines.  As long as the thermal energy exceeds the electrostatic counterpart, the 
  polarization is proportional to an external field and the dielectric constant is field independent.  But if the two energy scales 
  become comparable, polarization saturation takes place:  {the dipoles reach} perfect alignment with the field lines and no longer
  dissipate electrostatic energy above a certain threshold.  This behavior}  
  is described through the Langevin function  ${\mathcal L}(x)=\coth x-1/x$, and the polarization with nonlinear dependence 
  on the field strength is {$P = p_0 c_d {\mathcal L}(\beta p_0 |\bnabla\psi|)$, where $p_0$ is the dipole moment of a solvent {molecule} and 
  $c_d$ is the dipole density.  This} amounts to a change in dielectric constant according to
  \bea
    \epsilon_{\rm eff} \to \epsilon_0 + p_0 c_d \frac{{\mathcal L}(\beta p_0 |\bnabla\psi|)}{|\bnabla\psi|}.
  \eea 
  {Here, it is assumed that the solvent is incompressible and the modification  leads to the
  Langevin Poisson--Boltzmann equation  (LPB) \cite{Fry11}.}
  For small electrostatic field, $\epsilon \simeq \epsilon_0 + \beta p_0^2 c_d/3$ and 
  in the limit of large field $\epsilon\to\epsilon_0$.  {Uniform external fields affect only the dipole orientation. Nonuniform fields
  exert, in addition, a translational (dielectrophoretic) force, so that a dipole migrates toward regions with stronger fields.}  
  If one treats {the solvent as consisting of point dipoles}, as in Ref.~\cite{Abr07} (DPB, dipolar Poisson--Boltzmann equation), 
  this force leads to the 
  strong dependence of the solvent density on the field, $c_d(\beta p_0 |\bnabla\psi|)$,
  {with the functional form} $c_d(x)=\sinh x/x$.  
  {However, due to hydrogen bonds which lead to a high density of water at ambient conditions}, the assumption of 
  incompressibility is more realistic and the resulting $\epsilon $ qualitatively reproduces the drop with increasing field strength as 
  seen in the simulations \cite{Ber99}.

 \item {\em Ionic polarizabilities.} -- 
  {Ions can further be characterized by including {their} polarizability $\alpha$, so that {they} acquire an induced dipole moment in an
  external field.  Part of the electrostatic energy is dissipated when it polarizes ions.  This leads to a dielectric constant with 
  a term linear in the ion concentration,
  \bea
    \epsilon_{\rm eff} \to \epsilon + \alpha (c_+ + c_-).
  \eea   
  The presence of induced dipoles, in turn, contributes {to} a dielectrophoretic interaction of an ion with a field.  Accordingly, the 
  density distribution of ions becomes \cite{Fry11a},
  \bea
    c_{\pm} = c_s\exp(\mp e\beta\psi+\beta\alpha|\bnabla\psi|^2/2).  
  \eea     
  The above theoretical framework can be used to describe a somewhat different but related situation.  A dissolved charged 
  particle (for simplicity taken to be a point) is tightly bound to solvent dipoles forming {its} hydration shell {so that the dipoles
  become unresponsive to presumably a weaker external field.}  This produces a dielectric 
  cavity which reduces the dielectric constant of a solvent, on the one hand, and leads to a migration of a hydrated ion away from a
  strong field, on the other hand.  These trends are captured with polarizable model where the ``effective'' polarizability is a 
  negative valued quantity \cite{Hat12}.  Interestingly enough, both negative and positive $\alpha$ lead to a spreading out of a 
  diffuse double layer.  
  }

 \item {\em Stern layer.} --
  {A layer of solvent in direct contact with an interface has different properties from the bulk solvent.  
  First of all, ions cannot penetrate into it due to their finite size.  
  Second of all, the solvent within this layer tends to be more structured so that its dielectric response is expected to be 
  suppressed in relation to that in the bulk.  A frequent solution to incorporating this contribution is to introduce the Stern layer
  adjacent to a sharp  
  interface with the width corresponding to a radius of a bare or hydrated ion and the dielectric constant the same as 
  or lower than the bulk value.  The introduction of this layer does not change anything far from an interface, but it increases 
  the contact potential, $\psi_c \to \psi_c+\sigma_cR/\epsilon_{\rm stern}$, where $R$ is the radius of an ion and 
  $\epsilon_{\rm stern}$ is the dielectric constant with the Stern layer \cite{Fry11}.  Incorporation of the Stern layer at an air--water
  interface was found to give an accurate predictions for the excess surface tension of strong electrolytes \cite{Lev01}.}

\end{enumerate}

\begin{figure}[tbh]
\vspace{0.6cm}
\centerline{\resizebox{0.45\textwidth}{!}
{\includegraphics{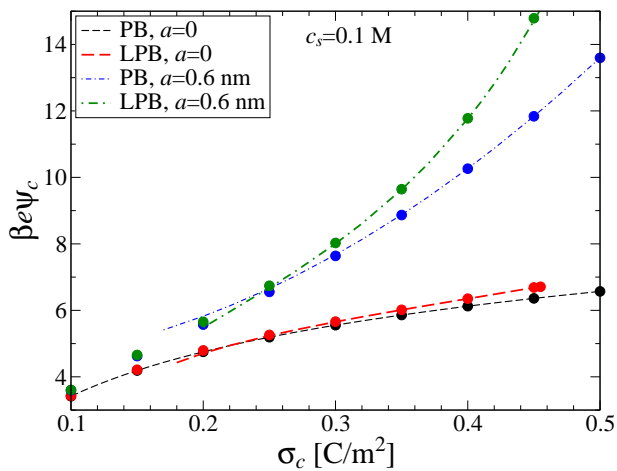}}}
 \caption{Contact potential (proportional to a colloid dipole moment at an interface) 
for PB and LPB, with and without steric
corrections due to finite ion size ($v=a^3$). The symbols correspond to numerical
data points and dashed lines correspond to analytical expressions 
derived in Ref.~\cite{Fry11}. Clearly, the contact potential in PB is noticeably smaller
than in the modified theories for higher charge densities, thus dipole moments will
be underestimated by PB alone.
}  
 \label{fig:phiwall}
\end{figure}

{As an example, in Fig.~\ref{fig:phiwall} we present a comparison between PB and MPB, LPB (routes (1) and (2) from above)
for the contact potential dependence at a charged wall on the charge density. Finite ion size and dielectric saturation lead to
a noticeable increase in the contact potential which is in line with the experimental results (discussed above) for the
dipole moment of charged colloids at interfaces.} 

\begin{figure}[tbh]
\vspace{0.6cm}
\centerline{\resizebox{0.45\textwidth}{!}
{\includegraphics{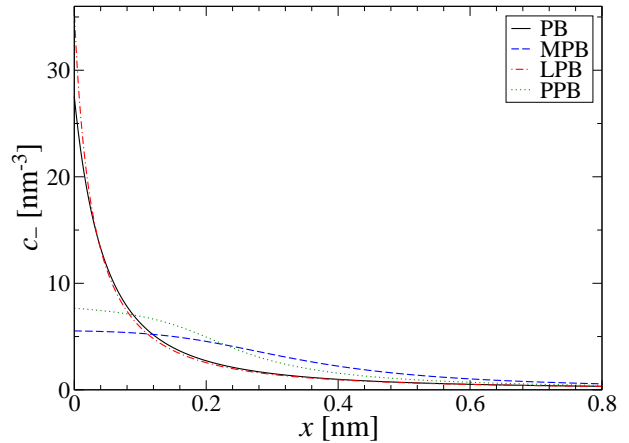}}}
 \caption{{Counterion density near a charged wall for the models discussed in Sec.~\ref{sec:mod_PB}.  
  The common parameters are: $\sigma_c=0.4\;{\rm C/m^2}$, $c_s=0.1\;{\rm M}$, and $\lambda_B=0.7$ nm.  Specific
  parameters for each model are: for 
  MPB the ion radius is $0.35{\rm nm}$, for LPB (point ions) the dipole concentration is $c_d=55{\rm M}$
  (as in water) and the permanent dipole moment is $p_0=4.85\;{\rm D}$, 
  for PPB the polarizability is $\alpha/(4\pi\epsilon_0)=-8\;{\rm M^{-1}}$.  Only LPB leads to an increased
  concentration near a wall, while the other contributions tend to spread out a diffuse layer.  }}  
 \label{fig:rho}
\end{figure}

{Fig.~\ref{fig:rho} compares counterion density profiles for various models.  The ``saturation'' effect is clearly seen for the
MPB model.  The mechanism behind the spreading-out of the diffuse layer for the PPB model is connected with the dielectrophoretic
repulsive force of ions with negative polarizability.  Only the LPB model causes a stronger attraction toward the wall.  
This is caused by the lowering of the dielectric constant close to a wall.  The combination of effects would require an appropriately 
combined model, an example is the combination of LPB and MPB as seen in the thick green dot--dashed curve in  Fig.~\ref{fig:phiwall}.  }

\section{Outlook to nonlocal descriptions}
 {In the models outlined above, the electrostatic interactions are described on the mean-field level, 
  and the {repulsive--core interactions among hydrated ions and solvent molecules} are incorporated as local contributions.  
  These models, even if quantitatively 
  not precise, are able to capture correct trends.  The drawback of the local approximation for the excluded volume 
  interactions is that the ``saturated'' density profile is an unphysical artifact of the method itself.  This
  leads to a {violation of the} contact value theorem (the relation between the density at the wall, {solvent pressure
  and wall potential}). 
  Despite this, the electrostatic properties at a charged 
  wall and far from the wall are captured with reasonable accuracy, despite the unphysical details in--between.  
  {A possibly more accurate treatment of the repulsive--core interactions can be achieved via 
  density functional theory. In the example of Ref.~\cite{Fry12}, ions are treated as charged hard spheres using
  the accurate fundamental measure theory, but for simplicity in the background of a structureless solvent. This corresponds
  to a nonlocal extension of the MPB modification discussed above.} {Finally, the mean-field treatment of electrostatics 
  neglects correlated fluctuations, which not only correct, but, in the so-called strong-coupling limit, 
  give rise to altogether new phenomena, such as the attraction between the same-charged macroparticles \cite{Naji05,Sam11}.  While the applicability 
  of the mean-field treatment breaks down in the strong-coupling limit, the correct theoretical treatment of 
  correlations continues to be a challenge.
  This challenge is increased if both the correlations due to the short-range 
  solvent and the long-range electrostatic interactions are important.  Currently, the solutions are looked for
  using either the density functional theory approach or the field-theoretical methods.  } 
  }

\begin{acknowledgments}
M.O. thanks the DFG (German Research Foundation) for support through
the Collaborative Research Center TR6 (Colloids in External Fields) and D.F. 
acknowledges support of CNPq (Brazil).
\end{acknowledgments}



\bibliography{chap.oettel.bib}      

\begin{thebibliography}{11}
\newcommand{\enquote}[1]{#1}
\providecommand{\natexlab}[1]{#1}
\providecommand{\url}[1]{\texttt{#1}}
\providecommand{\urlprefix}{URL }
\expandafter\ifx\csname urlstyle\endcsname\relax
  \providecommand{\doi}[1]{doi:\discretionary{}{}{}#1}\else
  \providecommand{\doi}{doi:\discretionary{}{}{}\begingroup
  \urlstyle{rm}\Url}\fi

\bibitem[{Abrashkin \emph{et~al.}(2007)Abrashkin, D. and Orland}]{Abr07}
Abrashkin, A., D., A. and Orland, H. (2007). \enquote{Dipolar poisson-boltzmann
  equation: Ions and dipoles close to charge interfaces,} \emph{Phys. Rev.
  Lett.} \textbf{99}, p. 077801.

\bibitem[{Bikerman(1942)}]{Bik42}
Bikerman, J.~J. (1942). \enquote{Structure and capacitance of electrical double
  layer,} \emph{Phil. Mag. Series 7} \textbf{33}, p. 384.

\bibitem[{Borukhov \emph{et~al.}(1997)Borukhov, Andelman and Orland}]{Bor97}
Borukhov, I., Andelman, D. and Orland, H. (1997). \enquote{Steric effects in
  electrolytes: A modified poisson-boltzmann equation,} \emph{Phys. Rev. Lett.}
  \textbf{79}, p. 435.

\bibitem[{Dominguez \emph{et~al.}(2008)Dominguez, Frydel and Oettel}]{Dom08}
Dominguez, A., Frydel, D. and Oettel, M. (2008). \enquote{Multipole expansion
  of the electrostatic interaction between charged colloids at interfaces,}
  \emph{Phys. Rev. E} \textbf{77}, p. 020401(R).

\bibitem[{Frydel(2011)}]{Fry11a}
Frydel, D. (2011). \enquote{Polarizable poisson--boltzmann equation: The study
  of polarizability effects on the structure of a double layer,} \emph{J. Chem.
  Phys.} \textbf{134}, p. 234704.

\bibitem[{Frydel \emph{et~al.}(2007)Frydel, Dietrich and Oettel}]{Fry07}
Frydel, D., Dietrich, S. and Oettel, M. (2007). \enquote{Charge renormalization
  for effective interactions of colloids at water interfaces,} \emph{Phys. Rev.
  Lett.} \textbf{99}, p. 118302.

\bibitem[{Frydel and Oettel(2011)}]{Fry11}
Frydel, D. and Oettel, M. (2011). \enquote{Charged particles at ﬂuid
  interfaces as a probe into structural details of a double layer,} \emph{Phys.
  Chem. Chem. Phys.} \textbf{13}, p. 4109.

\bibitem[{Frydel and Levin(2012)}]{Fry12}
Frydel, D. and Levin, Y. (2012). \enquote{A close look into the excluded volume 
  effects within a double layer,} \emph{J. Phys. Chem.} \textbf{137} p.164703.


\bibitem[{Hatlo \emph{et~al.}(2012)Hatlo, van Roij and Lue}]{Hat12}
Hatlo, M.~M., van Roij, R. and Lue, L. (2012). \enquote{The electric double
  layer at high surface potentials: The inﬂuence of excess ion
  polarizability,} \emph{EPL} \textbf{97}, p. 28010.

\bibitem[{Masschaele \emph{et~al.}(2010)Masschaele, Park, Furst, Fransaer and
  Vermant}]{Mas10}
Masschaele, K., Park, B.~J., Furst, E.~M., Fransaer, J. and Vermant, J. (2010).
  \enquote{Finite ion-size effects dominate the interaction between charged
  colloidal particles at an oil-water interface,} \emph{Phys.~Rev.~Lett.}
  \textbf{105}, p. 048303.

\bibitem[{Park \emph{et~al.}(2008)Park, Pantina, Furst, Oettel, Reynaert and
  Vermant}]{Par08}
Park, B.~J., Pantina, J.~P., Furst, E.~M., Oettel, M., Reynaert, S. and
  Vermant, J. (2008). \enquote{Direct measurements of the effects of salt and
  surfactant on interaction forces between colloidal particles at water-oil
  interfaces,} \emph{Langmuir} \textbf{24}, p. 1686.

\bibitem[{Yeh and Berkowitz(1999)}]{Ber99}
Yeh, I.~C. and Berkowitz, M.~L. (1999). \enquote{Dielectric constant of water
  at high electric ﬁelds: Molecular dynamics study,} \emph{J.~Chem.~Phys.}
  \textbf{110}, p. 7935.

\bibitem[{Levin and Flores-Mena(2001)}]{Lev01}
Levin, Y. and Flores-Mena, J.~E. (2001). \enquote{Surface tension of strong 
 electrolytes,} \emph{EPL} \textbf{56}, p. 187.

\bibitem[{Naji \emph{et~al.}(2005)Naji, Jungblut, Moreira, and Netz}]{Naji05}
Naji, A., Jungblut, S., Moreira, A. G. and Netz, R. R. (2005).
\enquote{Electrostatic interactions in strongly coupled soft matter,} \emph{Physica A} 
\textbf{352}, p. 131.


\bibitem[{Samaj and Trizac (2011)}]{Sam11} Samaj, L. and Trizac, E. (2011).  
\enquote{Counterions at highly charged interfaces: From one plate to like-charge attraction,} 
\emph{Phys. Rev. Lett.} \textbf{106}, p. 078301.


\end{thebibliography}


\end{document}